%% file: prb.tex
\documentclass[twocolumn,showpacs,prb,superscriptaddress,aps,floatfix]{revtex4-1}

\usepackage{rotating}
\usepackage{amsmath}
\usepackage{amssymb}
\usepackage{color}
\usepackage{graphicx}
\usepackage[active]{srcltx}
\usepackage[sort&compress]{natbib}
\usepackage{amssymb}
\usepackage[dvips]{epsfig}
\usepackage{float}

\include{alias}

\def\ai         {{\it ab-initio} }
\def\ga         {\alpha}
\def\gb         {\beta}

\def\gee        {\varepsilon}
\def\gl         {\lambda}

\def\go         {\omega}
\def\goql       {\omega_{\qq \gl}}
\def\gql        {\qq \gl}
\def\gS         {\Sigma}
\def\la         {\langle}
\def\ra         {\rangle}
\def\kk         {{\bf k}}
\def\qq         {{\bf q}}

\def\epi        {EP interaction }

\def\ep         {EP }

\def\nk         {n\kk}

\def\Znk        {Z_{\nk}}

\def\zbGN       {{\it zb}--GaN }

\def\zbGNv       {{\it zb}--GaN, }
\def\zbGNp       {{\it zb}--GaN. }

\renewcommand{\[}{\left[}
\renewcommand{\]}{\right]}
\renewcommand{\(}{\left(}
\renewcommand{\)}{\right)}

\newcommand{\grenoble}{Institut Laue Langevin BP 156 38042 Grenoble France}
\newcommand{\cnr} {Istituto di Struttura della Materia of the National Research Council, Via Salaria Km 29.3,
I-00016 Monterotondo Stazione, Italy}
\newcommand{\etsf} {European Theoretical Spectroscopy Facilities (ETSF)}
\newcommand{\tokyo} {Department of Chemical System Engineering, School of Engineering, The University of Tokyo, Tokyo 113-8656, Japan}

\begin{document}
\title{Electron--electron and electron--phonon correlation effects on the finite temperature electronic and optical properties of \zbGN}

\author{Hiroki Kawai}
\author{Koichi Yamashita}
\affiliation{\tokyo} 
\author{Elena Cannuccia}
\affiliation{\grenoble} 
\author{Andrea Marini}
\affiliation{\cnr} 
\affiliation{\etsf}

\date{\today}
\begin{abstract}
We combine the effect of the electron--electron and electron--phonon interactions to study the electronic and optical properties 
of \zbGNp 
We show that only by treating the two effects at the same time it is possible to obtain an unprecedented agreement of the zero and finite--temperature 
electronic gaps
and absorption spectra with the experimental results.
Compared to the state--of--the--art results our calculations predict a large effect on 
the main absorption peak position and width as well as on the overall absorption lineshape.
These important modifications are traced back to the combined electron--phonon damping
mechanism and non uniform $GW$ level corrections.
Our results demonstrate the importance of treating on equal footing the electron and phonon mediated correlation effects to obtain an accurate description of
the III--nitrides group physical properties.  
\end{abstract}           


\maketitle

\section{Introduction}
\label{intro}

The group III--nitride semiconductors, i.e., GaN, AlN, InN and their alloys are materials 
with many applications in the field of optoelectronics. These include, among others, light emitting diodes (LEDs), laser diodes (LDs), 
heterojuction field--effect transistors\,(HFETs)\cite{As2009,Novikov2008,Ambacher2002,Tschumak2010,Mietze2011}.
This class of compounds is widely used being characterized by the most stable wurtzite structure. They have
built--in electric fields arising from the spontaneous and piezoelectric polarization along the {\it c} axis. 
These fields are, however, undesirable in the applications of the heterostructures as 
quantum wells\,(QWs) or superlattices since they complicate the design and worsen the sample malleability. 
One of the approaches to eliminate these internal fields is the utilization of metastable non polar zinc--blende ({\it zb}) structures.
It has also been reported that {\it zb} group III--nitrides have a quantum confined Stark effect 
in low--dimensional heterostructures\cite{Feneberg2012}, high {\it p}--type conductivity in (Ga,Mn)N thin films\cite{Edmonds2005}
and negative differential resistance\,(NDR) at the resonant tunneling diode of the cubic Al(Ga)N/GaN\cite{Zainal2010, Mietze2012}.
Consequently a lot of interest is constantly attracted by this family of materials. 

In last few years \zbGN with high phase--purity and crystalline quality has been fabricated 
as a nearly strain--free epitaxial layer on 3C-SiC(001)/Si pseudo substrates
by plasma--assisted molecular beam epitaxy\cite{As2000,Feneberg2012,Mietze2012,As2013}.
This experimental achievement boosted the interest on fundamental optical properties as photoluminescence, 
photoreflectance and ellipsometry with particular attention on their temperature dependence.  

In contrast to such abundance of experimental results
the agreement with the state--of--the--art calculations of
the optical properties of \zbGN is still not satisfactory. 
In these approaches the absorption spectrum is calculated\cite{Benedict1999}
by including electron--hole interaction 
by solving the Bethe--Salpeter equation\,(BSE) derived within the Many--Body Perturbation Theory (MBPT)\cite{Onida2002}.
Nevertheless the main peak position is strongly underestimated when compared to the experimental result. And also the complex
temperature dependence observed experimentally is not captured at all.
Similarly, the band structure of \zbGN has been deeply investigated by using the most up--to--date theoretical approaches. 
In this case electron--electron correlation only has been included, by means of the well known $GW$ approximation\cite{gunnarson1998}.
The corresponding quasi--particle\,(QP) gap, calculated by using the one--shot $GW$ approximation on top of Kohn--Shame\,(KS) HSE hybrid orbital
\,(HSE+$G_0W_0$)\cite{Carvalho2011,Feneberg2012}, is $3.427$\,eV, which overestimates the experimental value of $3.295$\,eV\cite{Feneberg2012}.

The common denominator to these calculations of the electronic and optical properties is that electron--phonon\,(EP) interaction is
not considered. As a natural consequence no temperature dependence is captured. And, more importantly, also the well--known 
zero--point motion effect is neglected.
This assumption is, on the basis of very recent results~\cite{Giustino2010,cannuccia_2011,cannuccia_2013,ponce_2013,gonze}, not well--motivated.
Indeed the majority of the \ai\,\,simulations of the electronic and optical properties of
a wide class of materials are generally performed by keeping the atoms frozen in their crystallographic positions. 
Nevertheless, many years ago, Heine, Allen, and Cardona\,(HAC)\cite{allen1976,Cardona2006} pointed out the fact that the electronic states can be strongly affected 
by the lattice vibrations even when $T\rightarrow 0$\,K through the quantum zero--point motion effect. 
In the HAC approach the \ep interaction is treated in a static manner and the atomic displacements are considered as static perturbations.
The HAC approach successfully explained the temperature dependence of the gap shift and peak broadening in 
semiconductors like Si or Ge\cite{allen1983} .
BSE calculations on top of QP states including \ep correction have also been performed, 
showing remarkable \ep effect on the excitonic states and explaining the finite temperature 
evolution of the optical absorption measured experimentally\cite{marini2008}.

Despite of these successful results based on the HAC approach it has been recently 
discovered the key importance of considering dynamical corrections to the static HAC picture.
For instance, diamond has been shown to have large dynamical \ep effects, which explain
the subgap states observed experimentally in the absorption spectrum\cite{cannuccia_2011}.
Similarly, carbon polymer systems like \tpa and polyethylene, show a severe breakdown of the QP picture induced by the \ep
interaction\cite{cannuccia_2011,cannuccia_2013}.

In this work we calculate the electronic and optical properties of \zbGN by including electron--phonon and electron--electron interaction.
Our results show a remarkable impact of electron--phonon interaction even at zero--temperature which corrects
the overestimation of the QP gap obtained within the HSE+$G_0W_0$ method. 
At the same time we prove that only by treating on the same level electron--electron and  electron--phonon interactions it is possible to 
obtain and unprecedented agreement with experiment result, both at zero and finite temperature.

The paper is organized as following. 
In Sec.\ref{sec:MBPT} the \epi is briefly discussed in a MBPT framework.
In Sec.\ref{sec:Gap_renormalization} the electronic gap 
and transition energies at high--symmetry points of the Brilloun zone
of \zbGN are studied.
In Sec.\ref{sec:EP_BSE} we analyze the zero and finite temperaturs optical absorption by including both
electron--hole attraction and electron--phonon effects by using the BSE.

\section{A Many--Body perturbation theory approach to the electron--phonon problem}
\label{sec:MBPT}
The total Hamiltonian of the coupled electron--nuclei system $\hat{H}$ can be divided into three parts 
\begin{align}
\widehat{H} = \Hzero + \Hone + \Htwo, 
\label{eq:sec_theory_1}
\end{align}
where  $\hat{H_0}$ is the electronic Hamiltonian corresponding to the case where the atoms are frozen at their equilibrium positions $\RR_0$,
\begin{align}
\Hzero = \sum_i\[-\frac{1}{2} \frac{\partial^2}{\partial \hat{\rr_i}^2} +\left.\hat{V}_{ion}\[\{\RR\}\]\right|_{\RR=\RR_0}\(\rr_i\)\]+
 \hat{W}_{e-e}.
\label{eq:sec_theory_2}
\end{align}
$\Hone$ and $\Htwo$ represent, respectively, the first and second term in the Taylor expansion of $\Hzero$ 
when the atomic positions $\{\RR\}$ are expanded around the equilibrium positions $\{\RR_0\}$. At this stage
electron--electron correlations (described by $\hat{W}_{e-e}$) are treated at a mean--field level by using 
the standard Density--Functional Theory\.(DFT). In DFT $\Hzero\approx\sum_i\[\hat{h}\(\rr_i\)\]$ with
\begin{align}
\hat{h}\(\rr\) = -\frac{1}{2} \frac{\partial^2}{\partial \hat{\rr}^2} +\left.\hat{V}_{scf}\[\{\RR\}\]\right|_{\RR=\RR_0}\(\rr\),
\label{eq:sec_theory_2p}
\end{align}
and the 
derivatives of the
electronic effective potential $\Vscf = \widehat{V}_{ion} + \widehat{V}_{H} + \widehat{V}_{xc}$ with respect to the atomic coordinates $\RR$
can be calculated, self--consistently, by using  Density--Functional Perturbation Theory\,(DFPT). 

Within MBPT~\cite{cannuccia_2011,cannuccia_2013} the exact single particle excitation energies of the total Hamiltonian $\widehat{H}$ 
are obtained as poles of the Green's Function\cite{mattuck} $G_{\nk}(\go)$ that is solution
of the Dyson Equation:
\begin{align}
G_{\nk}\(\go\)=G^{\(0\)}_{\nk}(\go)\[1+\Sigma_{\nk}\(\go\)G_{\nk}\(\go\)\].
\label{eq:sec_theory_2a}
\end{align}
MBPT allows to calculate $\Sigma$ in terms of $\Hone$ and $\Htwo$.
We consider now the two lowest--order non--vanishing contributions to $\Sigma$ written as functionals of the non--interacting Green's function
$G^{0}_{\nk}(\go)$.
The second--order term in the perturbative expansion in powers of $\Hone$ gives the Fan contribution~\cite{fan1950} to the self--energy
\begin{multline}
\Sigma^{Fan}_{\nk}\(\go,T\) = \sum_{n'\gql} \frac {\gsq}{N_q} \times \\
\times\[ \frac{N_{\qq\gl}\(T\)+1-f_{n'\kk-\qq}}{\go-\gee_{n' \kk-\qq} -\goql -i0^{+}} \right. + \\
+ \left. \frac{N_{\qq\gl}\(T\)+f_{n' \kk-\qq}}{\go-\gee_{n' \kk-\qq}+\goql -i0^{+}}\],
\label{eq:sec_theory_8}
\end{multline}
where $\gee_{n' \kk-\qq}$ is Kohn--Sham energy of the $n'$th band at the point $\kk-\qq$ in the Brillouin zone.
$\goql$ is phonon energy relative to the mode $\lambda$ and transferred momentum $\qq$. 
$N_{\qq\gl}\(T\)$ is the Bose--Einstein distribution function of the phonon mode $\(\qq,\gl\)$ at temperature $T$ 
and $f_{n' \kk-\qq}$ is the occupation number of the bare electronic state at $\(n', \kk-\qq\)$.
$g^{\gql}_{n' n \kk}$ are the electron--phonon matrix element~\cite{cannuccia_2013} defined as:
\begin{multline}
g^{\gql}_{n n' \kk}=\sum_{s \ga} \(2 M_s \go_{\gql}\)^{-1/2} e^{i\qq\cdot\tau_s} \times \\
\times \la n\kk |\frac{\partial \Vscf\(\rr\)}{\partial{R_{s\ga}}} | n' \kk-\qq \ra \xi_{\ga}\(\qq \gl|s\),
\label{eq:gkkp}
\end{multline}
with $M_s$ the mass of the atom whose position in the unit cell is $\tau_s$. $\xi_{\ga}\(\qq \gl|s\)$ are the phonon polarization vectors.
As already pointed out all ingredients of Eq.(\ref{eq:gkkp}) are calculated by using DFPT.

Similarly to the Fan term, the Debye--Waller\,(DW) self--energy arises from the first--order term in the perturbative expansion in powers of $\Htwo$, 
\begin{align}
\gS^{DW}_{\nk}\(T\)=\frac{1}{N_q}\sum_{\gql} \Lambda^{\qq\gl,-\qq\gl}_{n n \kk}  \(2 N_{\qq\gl}\(T\) +1\), 
\label{eq:sec_theory_9}
\end{align}
where $\Lambda^{\qq\gl,-\qq\gl}_{n n \kk}$ is a second--order electron--phonon matrix element~\cite{cannuccia_2013}:
\begin{multline}
    \Lambda^{\gql,\qq'\gl'}_{n n' \kk}= \frac{1}{2}\sum_{s}\sum_{\ga,\gb} \frac{ \xi^{*}_{\ga}\(\qq \gl|s\) 
    \xi_{\gb}\(\qq' \gl'|s\)}
{2M_s\(\go_{\gql} \go_{\qq' \gl'} \)^{1/2}} \times \\
\times \la n\kk |\frac{\partial^2 \Vscf\(\rr\)}{\partial{R_{s\ga}}\partial{R_{s\gb}}} | n' \kk-\qq-\qq' \ra.
\label{eq:lambda_factors}
\end{multline}

By solving explicitly Eq.(\ref{eq:sec_theory_2a}) 
the fully interacting Green's function $G_{nk}\(\go,T\)$ can be written as 
\begin{equation}
G_{\nk}\(\go,T\)=\frac {1}{\go-\gee_{\nk}-\gS^{Fan}_{\nk}\(\go,T\)-\gS^{DW}_{\nk}\(T\)}.
\label{eq:Dyson}
\end{equation}
The imaginary part of the Green's function $A_{\nk}\(\go,T\)\equiv\pi^{-1}\mid\Im\[G_{\nk}\(\go,T\)\]\mid$
gives the electronic spectral function\,(SF). 
In the quasi--particle approximation\,(QPA) the SF is assumed to be well described by a lorentzian function.
Mathematically this 
means that the self--energy frequency dependence can be expanded 
linearly around the bare electronic energy. In this case, the pole of  $G_{nk}\(\go,T\)$, $E_{\nk}(T)$ is given by
\begin{multline}
E_{\nk}\(T\) = \gee_{\nk} + \\+ Z_{\nk}\(T\) \[\gS^{Fan}_{\nk}\(\gee_{\nk},T\)+\Sigma^{DW}_{\nk}\(T\)\],
\label{eq:QP_energy}
\end{multline}
with $Z_{\nk}\(T\) = \(1-\left. \frac{\partial \gS^{Fan}_{\nk}\(\go,T\)}{\partial \go}\right|_{\go=\gee_{\nk}}\)^{-1}$ representing the
renormalization factor.  
The on--the--mass--shell\,(OMS) approximation represents the static limit of the QPA, obtained by assuming 
$\gS^{Fan}_{\nk}\(\go,T\)\approx \left.\gS^{Fan}_{\nk}\(\go,T\)\right|_{\go=\gee_{\nk}}$,
which is equivalent to assume $Z_{\nk}\(T\)=1$ in the QPA. 
The Fan and DW self--energies are complex and real functions, respectively,  
thus the former gives both an \ep induced energy shift and broadening while the latter contributes only with a constant energy shift. 
Both self--energies depend explicitly on the temperature $T$ via the $N_{\qq\gl}\(T\)$ factor.

\section{Renormalization of the single particle energy levels. 
The combined effect of the electron--electron and electron--phonon interactions}
\label{sec:Gap_renormalization}
\zbGN  is a polar material and, as a consequence, large static \ep effects are expected~\cite{Botti2013}.
As mentioned above a strong \ep coupling can eventually induce the breakdown of the QPA. A clear and simple way to test the QPA validity is
to calculate the renormalization factors $\Znk\(T\)$. Indeed, by using Eq.(\ref{eq:Dyson}) and Eq.(\ref{eq:QP_energy}) it turns out that,
within the QPA, the Green's function $G_{\nk}^{QP}\(\go,T\)$ can be written as
\begin{align}
G_{\nk}^{QP}\(\go,T\)= \frac {\Znk\(T\)}{\go-E_{\nk}\(T\)},
\label{eq:Dyson_QP}
\end{align}
with $E_{\nk}\(T\)$ evaluated by means of Eq.(\ref{eq:QP_energy}).
When $\Znk=1$ the SF, $\Im|G_{\nk}^{QP}\(\go,T\)|$, reduces to a lorentzian function
with a pole at $\go=\Re\[E_{\nk}\(T\)\]$ and width $\Gamma_{\nk}\(T\)=\Im\[E_{\nk}\(T\)\]$.
Thus, the $\Znk$ values measure the strength of the quasi--particle pole, i.e.,
the QP picture is well motivated when the SF can be well approximated with a single Lorentzian--like function.

\begin{figure}[H]
\epsfig{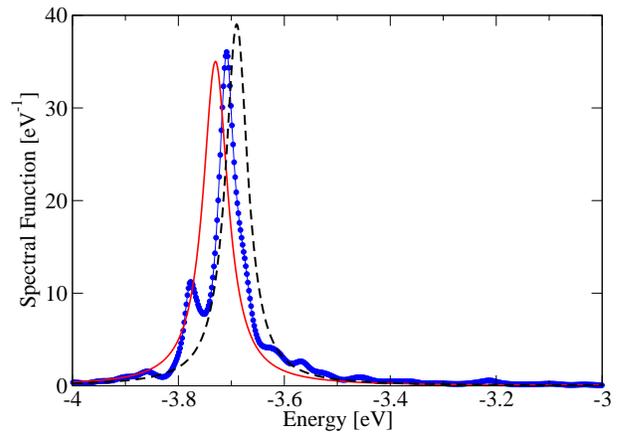}
\caption{\footnotesize{Spectral function of a valence band state. This well represents the general trend of 
the majority of spectral functions covering the energy range involved in the absorption process, as discussed in the text.
Blue line with dots is the calculated SF. This is compared with two Lorentzian functions corresponding to the OMS (red solid line) and to the 
QPA (black dash line). Both approximation reproduce the calculated SF quite well and the use of the OMS is, therefore, well motivated.
}}
\label{fig:SF_52}
\end{figure}

In our \ep calculations the optimized geometry and the electronic state are obtained by using the PWSCF code\cite{pwscf}.
Electron--phonon calculations are performed with the Yambo code\cite{Marini20091392} by using the 
phonons frequencies and $g^{\gql}_{n n' \kk}$ matrix elements calculated with PWSCF within DFPT. 
As a results of our simulations the majority of the states that contribute to the optical absorption are well
described by Lorentzian--like SF, as shown in Fig.\ref{fig:SF_52}. In addition most of the states 
show values of $\Znk$ very close to 1.
For example, the states corresponding to the valence band maximum\,(VBM) and the conduction band minimum\,(CBM) at the $\Gamma$ point,
have $\Znk=0.91$ and $\Znk=0.98$, respectively. 

This indicates that, in \zbGNv the OMS approximation is well motivated and most of the weight can be safely assumed to be in one single peak. 
There is, however, another and more stringent motivation in favor of OMS as far as the calculation of the optical properties is concerned. 
A $\Znk$ factor smaller than 1 is known to reduce the
intensity of the absorption spectrum. At the same time, however, it is well known that such reduction is compensated by the dynamical electron--hole 
interactions~\cite{DelSole1996,marini2003}. As far as these dynamical effects are neglected (as commonly done in the state--of--the--art
implementation of the BSE used in this work) the OMS assumption of $\Znk=1$ is well motivated also from  a purely theoretical point of view.

In order to describe the impact of \ep on the electronic states we consider the energies corresponding to the
lowest transition energies at several high--symmetry points. These energies are compared with the 
experimental results in Tab.~\ref{tab:VB_CB}.
DFT is well known to underestimate the band gaps of about $\simeq{40\%}$.
In fact, our DFT calculation (performed with the LDA) yields $2.231$\,eV as the band gap of \zbGNp This is
clearly less than the experimental value, that is $3.295$\,eV at 10 K\cite{Feneberg2012}.
Our LDA+$G_0W_0$ calculations within the plasmon--pole approximation~\cite{gunnarson1998}
opens the gap to $3.239$\,eV, which well agrees with the experiment.
Still, the combination LDA+$G_0W_0$ largely underestimates the  transition energies
at L and X.

This underestimation can be traced back to the local treatment of
electron--electron correlation effects in the self--consistent DFT--LDA calculation. 
This limitation can be overcome by using the AM05~\cite{PhysRevB.72.085108} approximation 
for the exchange--correlation energy functional to calculate the optimized geometry
and the HSE\cite{heyd:8207} functional for the start point of $G_0W_0$ calculation. 

From Tab.~\ref{tab:VB_CB} it is evident that HSE+$G_0W_0$ overestimates the transition energies
at all three high symmetry points\cite{Carvalho2011}. At the same time, however,
the \ep interaction greatly reduces this overestimation 
leading to an excellent agreement with the experimental results.
Our calculation at OMS level gives a gap correction  of
$-0.127$ eV at $\Gamma$, which reduces 
the HSE+$G_0W_0$ gap to $3.300$\,eV, in agreement with the experiment.
Similarly, at the L and X points, the \ep induced correction is $-0.190$\,eV and $-0.132$\,eV resulting in transition energies of $7.517$\,eV and
$7.624$\, eV, again in very good agreement with the experiment.
This result indicates the importance of the \ep correction in \zbGNv pointing to similar and potentially important corrections in the whole
III--nitrides group of materials. 

\begin{table}[h]
\caption{\footnotesize{
Lowest transition energies at high--symmetry points in the Brillouin zone of \zbGNp The values obtained 
from LDA, LDA+$G_0W_0$, HSE+$G_0W_0$\cite{Carvalho2011} and HSE+$G_0W_0$+OMS calculations are compared with the experimental
values\cite{Feneberg2012}.
All values are in eV.
}}
\setlength{\tabcolsep}{12pt}
\begin{tabular}{cccc}   \hline \hline
                           &  $ \Gamma$   &   $ \rm{L}$        &  $ \rm{X}      $    \\ \hline 
  LDA                      &   $ 2.231 $  &   $ 5.952 $        &  $ 6.034       $    \\ 
  LDA+$G_0W_0$             &   $ 3.239 $  &   $ 7.117 $        &  $ 7.105       $    \\ 
  HSE+$G_0W_0$             &   $ 3.427 $  &   $ 7.707 $        &  $ 7.755       $    \\ 
   HSE+$G_0W_0$+OMS        &   $ 3.300 $  &   $ 7.517 $  &  $ 7.624 $    \\ \hline 
  Exp ($T=10$ K)           &   $ 3.295 $  &   $ 7.33{\rm ^a}$  &  $ 7.62{\rm ^a}$    \\ \hline \hline
\end{tabular}
{\footnotesize{
\flushleft{
$^a$Excitation peak energies.
}
}}
\label{tab:VB_CB}
\end{table}

\section{Finite temperature optical absorption spectra including electron--hole effects}
\label{sec:EP_BSE}
The optical absorption spectrum is defined as the imaginary part of
the macroscopic dielectric function $\Im\[\epsilon_M\(\go\)\]$. This can  be easily expressed,
in the long wave length limit, as 
\begin{multline}
\epsilon_M\(\go\)=\\
1-\lim_{\qq\to0}v_0\(\qq\)\int d\rr d\rr' e^{-i\qq(\rr-\rr')} \overline\chi\(\rr,\rr';\go\), 
\label{eq:macroscopic_epsilon}
\end{multline}
with $v_\GG\(\qq\)=4\pi/|\GG+\qq|^2$ the Coulomb potential and $\overline\chi\(\rr;\rr';\go\)$ the two--point polarizability.
The equation of motion for the polarizability\cite{Onida2002} can be rewritten by
introducing a single--particle basis set ($\{\phi_{n,\kk}\}$) to expand the density operator. This is equivalent
to define the electron--hole probability functions 
$\Phi_\KK\(\rr\)=\phi_{c\kk}\(\rr\)\phi^*_{v\kk}\(\rr\)$. Here $\KK$ represents the general
conduction--valence pairs, $\KK=\(c,v,\kk\)$. In this basis $\overline\chi$ is
\begin{multline}
\overline\chi\(\rr;\rr';\go\)=\\
-\(\frac{i}{\Omega N}\)\sum_{\KK_1,\KK_2}\Phi^{*}_{\KK_1}\(\rr\)L_{\KK_{1}\KK_{2}}\(\go\)\Phi_{\KK_2}\(\rr'\).
\label{eq:fourpoint_expanded}
\end{multline}
Eq.(\ref{eq:fourpoint_expanded}) introduces the electron--hole Green's function $L_{\KK_{1}\KK_{2}}\(\go\)$ that
satisfies the BSE equation~\cite{Onida2002}
\begin{multline}
L_{\KK_{1}\KK_{2}}\(\go\)=\\
L^{0}_{\KK_{1}\KK_{2}}\(\go\)+L^{0}_{\KK_{1}\KK_{3}}\(\go\)
\Xi_{\KK_{3}\KK_{4}}\(\go\)L_{\KK_{4}\KK_{2}}\(\go\).
\label{eq:BSE}
\end{multline}
The Bethe--Salpeter kernel $\Xi$ is defined as $\Xi=-iV+iW$ with
$V$ and $W$ the exchange and screened Coulomb interactions, respectively.
$L^{0}_{\KK_{1}\KK_{2}}\(\go\)$, in Eq.(\ref{eq:BSE}), is the free electron--hole Green's function, defined in 
Eq.(\ref{eq:L0_temp}).

As previously described by Marini\cite{marini2008} it is possible to include finite--temperature effect in the BSE
by using, as reference single--particle energies, the temperature--dependent and complex QP energies $E_{\nk}\(T\)$. In this way
the free electron--hole Green's function $L^{0}_{\KK_{1}\KK_{2}}\(\go, T\)$ depends explicitly on the temperature 
\begin{multline}
L^{0}_{\KK_{1}\KK_{2}}\(\go, T\)=
\\i\[\frac{f_{c_1\kk_1}-f_{v_1\kk_1}}{\go-E_{c_1\kk_1}\(T\)+E_{v_1\kk_1}\(T\)+i0^{+}}\]\delta_{\KK_{1}\KK_{2}}. 
\label{eq:L0_temp}
\end{multline}
Eq.(\ref{eq:L0_temp}) ensures that also the fully interacting electron--hole Green's function
and the absorption spectra depend explicitly on the temperature, thanks to Eq.(\ref{eq:macroscopic_epsilon}) and Eq.(\ref{eq:fourpoint_expanded}).

In order to solve the BSE we adopt two standard approximations. The first is the
Tamm--Dancoff approximation which corresponds to neglect the coupling between the resonant and the anti--resonant part of the
BSE kernel. The second is the use of a statically screened electron--hole potential $W$.
 
\begin{figure}[t]
\epsfig{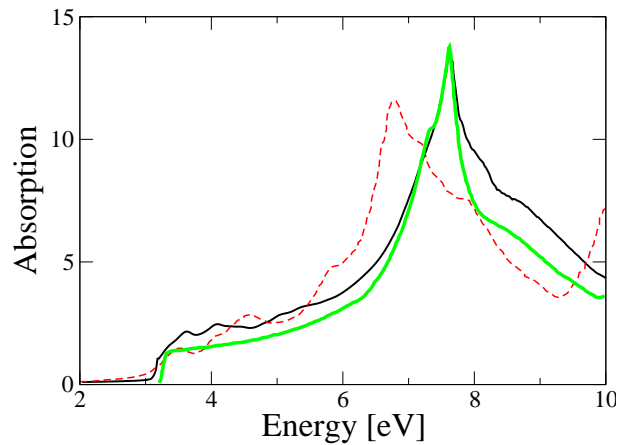}
\caption{\footnotesize{ 
Theoretical and experimental absorption spectra of \zbGN at $T=10\,K$.
The spectrum obtained by solving the BSE including the \ep correction (black thin line)
is in excellent agreement with the experimental result\cite{Feneberg2012} (green bold line).
Compared to the state--of--the--art calculation of  Benedict {\it et al.},\cite{Benedict1999} (dashed red line) the agreement is largely improved.
}}
\label{fig:BSE_121212_GW}
\end{figure}

In Fig.\ref{fig:BSE_121212_GW} we show the calculated absorption spectrum.
In addition to the $G_0W_0$ corrections calculated as described in Sec.\ref{sec:Gap_renormalization} we 
include \ep effects. 
To obtain a converged absorption spectra
we employed the random--integration method (RIM)\cite{Marini20091392} by selecting around 30000 random $\kk$-points in the whole Brillouin Zone.
The resulting spectrum (thick black line) is compared to the previous calculation of 
Benedict {\it et al.}\cite{Benedict1999} (dashed red line) which is performed in a  
LDA basis without including the \ep interaction. 
The present calculation, instead, is in excellent agreement with the $T=10K$ experimental spectrum\cite{Feneberg2012}\,(green bold line)
and largely improves the Benedict result.
We notice, indeed, that, compared to the Benedict calculation, the position and width of the main peak is in very good agreement with the
experiment. 
This pronounced peak located at around $7.62$\,eV is due to
interband transitions in a region of the Brillouin Zone near the X point. These transitions extend to regions 
where the valence and conduction bands are parallel with a similar energy distance. 

It is crucial to underline that, in the present work, the absorption spectrum width is dictated by the \epi and it is temperature--dependent.
Indeed we only use a very small artificial damping (10\,meV) 
in Eq.(\ref{eq:L0_temp}) to avoid numerical instabilities. Compared to the case of Benedict, where this broadening
is arbitrarily chosen to be  200\,meV, the \ep interaction correctly describes both the main peak
and the steep absorption edge. 

\begin{figure}[t]
\epsfig{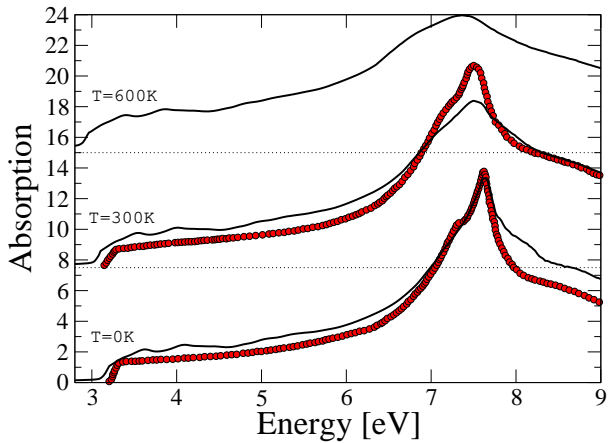}
\caption{\footnotesize{
Absorption spectra of \zbGN at $T=$ 0, 300, and 600 K.
Red circles are experimental results at 10 K and 295 K.\cite{Feneberg2012}
}}
\label{fig:finiteTemp}
\end{figure}

Finally we investigate how the optical spectrum evolves as the temperature is increased.
In Fig. \ref{fig:finiteTemp} we show the calculated absorption spectra at $T=0$\,K, $300$\,K and $600$\,K.
In the first two cases the numerical simulation is compared with the available experimental results~\cite{Feneberg2012}.
The agreement is fairly good and confirms that the present approach is able to correctly capture the finite temperature effects.
The energy shifts of the first excitation peaks due to electron--hole transitions occurring at $\Gamma$, L and X is experimentally
-63 meV, -100 meV and -110 meV  when the temperature is
increased from $10$\,K to $295$\,K\cite{Feneberg2012}.
Our calculations give -83 meV, -144 meV, and -106 meV, respectively, which show a quntitative agreement with experiment. 
Also the general trend observed experimentally that the peak shift is larger at critical points with higher transition energy is reproduced. 

On the other hand, from Fig.\ref{fig:finiteTemp} we deduce that the broadening of the main peak at the X point is slightly overestimated at
$T=300$\,K compared to the experiment. In order to understand the source of this overestimation we notice that,
in the QP picture, the \ep induced broadening of the valence band top and conduction band bottom at the X point are 121.8 meV and 12.1 meV at $T=300$ K.  
In the independent particle approximation (where  $L_{\KK_{1}\KK_{2}}\approx L^{0}_{\KK_{1}\KK_{2}} \delta_{\KK_{1}\KK_{2}}$)
the electron--hole broadening is simply the sum of the two.  Now, as also in the case where electron--hole attraction is included,
the main absorption peak originates from transitions concentrated around the VBM and CBM we deduce that the overestimation is due to a too large 
broadening of the underlying QP states.

In the experimental work by Logothetidis {\it et al.}\cite{Logothetidis1994} the broadening at the main absorption peak is described by a 
phenomenological model
\begin{align}
\Gamma\(T\) = \Gamma_1 + \Gamma_0\left[1+\frac{2}{\exp(\Theta/T)-1}\right], 
\label{eq:width_fit}
\end{align}
with $\Gamma_1=27$ meV,  $\Gamma_0 = 44$ meV, and $\Theta = 522$\,K. 
The first term in $\Gamma_1$ is to describe temperature--independent mechanism, as surface scattering, 
thus we set it to 0 to compare Eq.(\ref{eq:width_fit}) with our theoretical results. 
Eq.(\ref{eq:width_fit}), indeed, predicts  $\Gamma\(T=300\)\sim 62.7$ meV that is half of the value that results from the solution of the BSE.

A reasonable explanation of this deviation is in the underlying unperturbed band structure.
The band curvature has a large impact on the \ep induced broadening 
through the denominator of Eq.(\ref{eq:sec_theory_8}),
especially by the dominant intraband--scattering terms with $\go=\gee_{n \kk}$, $n'=n$, and small $\qq$. 
Since our \ep self--energies are calculated on top of Kohn--Sham states from LDA,
the resulted band widths are too small.
As shown in previous calculations~\cite{Carvalho2011} the valence band at the X point is characterized by a large curvature that is
underestimated by the LDA calculations.
We expect that the broadening would be improved by an \ep calculations performed on top of HSE+$G_0W_0$ band structure, 
but it is prohibitively expensive from the computational point of view.
Nevertheless our approach, based on LDA, gives excellent results especially at the low temperature. 

\section{Conclusion}
\label{sec:conclusion}
In this work we study the zero and finite temperature electronic and optical properties of \zbGNp
The effect of \epiv treated in a fully dynamical approach based on the Many--Body Perturbation Theory,
shows that the simple on--the--mass--shell approximation to the quasi--particle energies and widths is well motivated for
the low energy states involved in the absorption spectrum.

By including, in an \ai manner, the combined effect of the electron--electron and the \epi 
we obtain an excellent agreement with the experimental foundamental band gaps.

The solution of the BSE calculated on top of the HSE+$G_0W_0$ band structure including \ep effects
leads to an excellent agreement also for the optical absorption spectrum measured on high phase--purity samples. 
Both the position and the broadening of the most intense absorption peak
are correctly reproduced in the low--temperature regime.
In the room--temperature case, instead, the red--shift of the main peak position is well described while
the broadening is sliglhtly overestimated. 
Despite this overestimation the present results
still represent a major improvement with respect to the state--of--the--art 
simulations.

Our results clearly point to the crucial important of  including at the same time electron--electron and electron--phonon correlation effects
for a comprehensive and quantitative  understanding 
of the electronic and optical properties of group III--nitrides.

\section*{Acknowledgments}
H. K is supported by JSPS Research Fellowships for Young Scientists and JSPS KAKENHI (24-7666). 
A. M. acknowledges funding by MIUR FIRB Grant No. RBFR12SW0J.

\bibliographystyle{h-physrev}
\bibliography{prb}	

\end{document}

%% file: alias.tex
\newcommand{\be}{\begin{equation}}
\newcommand{\ee}{\end{equation}}
\newcommand{\bea}{\begin{eqnarray}}
\newcommand{\eea}{\end{eqnarray}}

\def\rr		{{\bf r}}

\def\GG		{{\bf G}}

\def\RR		{{\bf R}}
\def\KK		{{\bf K}}

\def\qq		{{\bf q}}
\def\kk		{{\bf k}}

\def\ga         {\alpha}
\def\gb         {\beta}

\def\gee        {\epsilon}
\def\gl         {\lambda}

\def\go         {\omega}

\def\goql       {\omega_{\qq \gl}}
\def\gql        {\qq \gl}

\def\gS         {\Sigma}

\def\la         {\langle}
\def\ra         {\rangle}

\renewcommand{\[}{\left[}
\renewcommand{\]}{\right]}
\renewcommand{\(}{\left(}
\renewcommand{\)}{\right)}

\def\tpa        {{\it trans}--polyacetylene }

\def\Vscf       {\widehat{V}_{scf} }
\def\Hzero      {\widehat{H}_0}
\def\Hone       {\widehat{H}_1}
\def\Htwo       {\widehat{H}_2}

\def\epi        {electron--phonon interaction }

\def\epiv        {electron--phonon interaction,}
\def\ep         {electron--phonon }

\def\nk         {n{\bf k}}

\def\gsq         {{\mid g^{\gql}_{n n' \kk} \mid}^2}

\def\bverb      {\renewcommand{\baselinestretch}{1}\begin{verbatim}}
\def\everb  	{\end{verbatim}\renewcommand{\baselinestretch}{1.3}}